\documentclass[prl,twocolumn,showpacs,superscriptaddress,amsmath,amssymb,amsfonts]{revtex4-1}
\usepackage{textcomp}
\usepackage{graphicx}
\usepackage{dcolumn}
\usepackage{bm}

\begin{document}

\title{A Triplet Resonance in Superconducting Fe$_{1.03}$Se$_{0.4}$Te$_{0.6}$}

\author{Juanjuan Liu}
\affiliation{Department of Physics, Renmin University of China, Beijing 100872, China}
\author{A. T. Savici}
\author{G. E. Granroth}
\affiliation{Neutron Scattering Division, 
Oak Ridge National Laboratory, Oak Ridge, Tennessee 37831, USA}
\author{K. Habicht}
\affiliation{Helmholtz-Zentrum Berlin f\"{u}r Materialen und Energy, D-14109 Berlin, Germany}
\author{Y. Qiu}
\affiliation{NIST Center for Neutron Research, National Institute of Standards
and Technology, Gaithersburg, MD 20899, USA} 
\affiliation{Department of Materials Science and Engineering, University of Maryland, College Park, MD 20742, USA}
\author{Jin Hu}
\affiliation{Department of Physics,  University of Arkansas, Fayetteville, AR 72701, USA}
\author{Z.Q. Mao} 
\affiliation{Department of Physics, Pennsylvania State University, University Park, PA 16802, USA}
\author{Wei Bao}
\email{wbao@ruc.edu.cn}
\affiliation{Department of Physics, Renmin University of China, Beijing 100872, China}

\date{\today}

\begin{abstract}
From heavy fermion compounds and cuprates to iron pnictides and chalcogenides, a spin resonance at $\hbar\Omega_0\propto k_BT_c$ is a staple of nearly magnetic superconductors. Possible explanations include a two-particle bound state or loss of magnon damping in the superconducting state. While both scenarios suggest a central role for magnetic fluctuations, distinguishing them is important to identify the right theoretical framework to understand these types of unconventional superconductors. Using inelastic neutron scattering technique, we show that the spin resonance in the optimally doped Fe$_{1.03}$Se$_{0.4}$Te$_{0.6}$ superconductor splits into three peaks in a high magnetic field, a signature of a two-particle $S=1$ triplet bound state. 
\end{abstract}


\maketitle

Initially discovered in cuprates \cite{ybco_resn,ybco_pol,BSCCO_resn,TBCO_resn,jmt_resn} and heavy-fermion \cite{U123_resn,U123_2001,Ce122_resn,Co115_stock} superconductors, and then in iron pnictide/chalcogenide superconductors \cite{A073932,A114755,B053559}, a spin resonance is arguably the most prominent experimental feature of unconventional superconductors \cite{UemuraII}. The leading interpretation of the resonance is an $S=1$ triplet excited bound state formed by two Bogoliubov-de Gennes quasiparticles in the superconducting state. The resonance intensity is enhanced by the coherence factor of a momentum-space modulated superconducting order parameter \cite{resn_1,resn_2,resn_3,resn_t_3,A032740,A041793,Co115_theory}. A different class of theories interpret
the resonance as a remnant magnon of the parent antiferromagnetic state \cite{dkm_sw,cdb_sw}. The intensity enhancement below the superconducting transition temperature $T_c$ is attributed to a
loss of the particle-hole damping channel inside the superconducting gap. While the two interpretations may have merit in different materials, we report here an experimental result that makes a clear distinction for the optimally doped $\rm Fe(Se,Te)$ superconductor. By splitting the resonance into three modes in a high magnetic field, we show that it is associated with a spin $S=1$ triplet, thus pointing decidedly to a two-particle bound state. 
In contrast, a conventional magnon is generally
a doublet that may split into two modes in a field \cite{spin_wave}.

The Fe$_{1+y}$Se superconductor \cite{A072369} contains the same electronically active iron-chalcogen or pnictogen layers as in 
other families of the iron-based superconductors \cite{Kamihara2008,A054630,A064688}.
The small amount of interstitial Fe  \cite{A092058} is necessary to stabilize the tetragonal $P4/nmm$ crystalline structure \cite{A072369}.
The superconducting transition temperature $T_c$ can be raised from 8 K to 14.6 K
through partial substitution of Se by Te \cite{B040824}, and the superconducting volume fraction can be enhanced and the transition sharpened by tuning the iron/chalcogen ratio close to one
($y\approx 0.03$) \cite{C035647}. A fortuitous aspect of the Fe(Se,Te) superconductors is the availability of large high-quality single crystalline samples, which allowed us to
unambiguously observe the development of a spin resonance above a gap in the superconducting state of Fe$_{1.03}$Se$_{0.4}$Te$_{0.6}$ in inelastic neutron scattering experiments \cite{B053559}. As in optimally doped BaFe$_{1.84}$Co$_{0.16}$As$_2$ ($T_c=22$ K) \cite{A114755}, the resonance of Fe$_{1.03}$Se$_{0.4}$Te$_{0.6}$ disperses only with in-plane momenta although its intensity varies along various directions \cite{B053559,B114713,D095196}, indicating a quasi-two-dimensional superconducting state. The spin-space of the resonance is anisotropic with the in-plane component $\sim 30\%$ larger than that along the $c$-axis \cite{PRB_pol_ILL}.

Single crystals of Fe$_{1.03}$Se$_{0.4}$Te$_{0.6}$ were grown by a flux method \cite{B040824}. 
An optimized set of Fe$_{1.03}$Se$_{0.4}$Te$_{0.6}$ single crystals
with narrower mosaic and higher $T_c =14.6$~K  than those used previously \cite{B053559} was prepared for this experiment. 
Three crystals, weighing 7.9, 3.8 and 3.6 g respectively, were mutually aligned so that the reciprocal ($h,k,0$) plane corresponds to
the horizontal scattering plane of the V2 FLEX cold neutron triple-axis spectrometer at HZB.
Since this experiment at high magnetic field would be severely intensity-limited, 
in addition to using the largest amount of samples possible within the magnet core, 
this sample orientation allows for integration of the quasi-two-dimensional resonance intensity along the $c$-axis \cite{B053559} by the vertical focusing of the spectrometer.
Neutron scattering measurements were conducted using the fixed $E_f$=5 meV configuration with a cold Be filter placed after the sample, and pyrolytic graphite was used for both monochromator and analyzer. 
The lattice parameters of the tetragonal $P4/nmm$ unit cell are $a=b=3.802$ and
$c=6.061 \rm{\AA}$ at room temperature. The mosaic full width at half maximum of the individual samples ranges from 1.2 to 1.6\textdegree, and the co-aligned assembly mosaic was 2.2\textdegree. The sample temperature 
and applied magnetic field were controlled by a 15 T vertical-field cryomagnet. 
The magnetic field was applied along the $c$-axis, thus perpendicular to the Fe square plane. 

Figure~1 shows constant {\bf q}=($\frac{1}{2},\frac{1}{2},0$) scans at temperature $T=1.7$~K. 
The upturn at high energies in all data sets is associated with the low scattering angle limit of the spectrometer where the incident beam impinges on the detection system.
In zero magnetic field, the resonance is centered at $\hbar\Omega_0= 7.1(1)$ meV. The ratio $\hbar\Omega_0/k_BT_c\approx5.6$ is comparable to the previously reported value for Fe(Se,Te)\cite{B053559} and cuprates YBa$_2$Cu$_3$O$_{6+x}$ \cite{ybco_resn}, Bi$_2$Sr$_2$CaCu$_2$O$_{8+x}$ \cite{BSCCO_resn} and Tl$_2$Ba$_2$CuO$_{6+x}$\cite{TBCO_resn}. However, it is larger than 4.3-4.5 reported for doped BaFe$_2$As$_2$ \cite{A073932,A114755} and 2-4 reported for La$_{2-x}$Sr$_x$CuO$_4$ \cite{jmt_resn} and heavy fermion superconductors \cite{U123_resn,U123_2001,Ce122_resn,Co115_stock}. 
In a 14 T magnetic field, the peak splits and fine structure appears. In particular, there are two clearly defined maxima at $\sim$5  and 7 meV, while the field dependence above 8.5 meV is less than the statistical accuracy in this background dominated part of the spectrum. 
\begin{figure}

\includegraphics[width=86mm,angle=0]{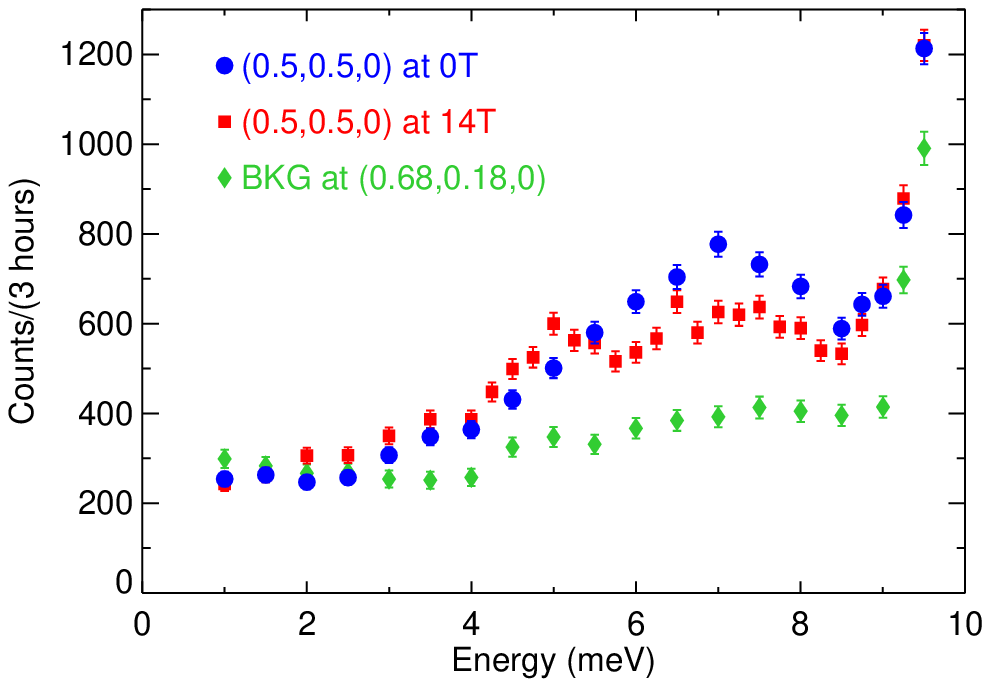}

\caption{Low temperature (1.7 K) inelastic neutron scattering data from $\rm Fe_{1.03}Se_{0.4}Te_{0.6}$ in the superconducting state. Blue circles are zero field data, while red squares were acquired in a field of 14 T. Green diamonds represent the background, acquired with the sample rotated 30\textdegree~from {\bf q}=(0.5,0.5,0), corresponding to {\bf Q}=(0.68,0.18,0).  }
\label{fig1}
\end{figure}

To clearly extract the magnetic signal, especially at higher energies, we measured a background spectrum at a wave vector transfer {\bf Q}=(0.68,0.18,0) where magnetic scattering was negligible \cite{B053559}. The background subtracted data were normalized to represent the differential scattering 
cross-section $S(\bf{q},\omega)$ by comparing to 
energy integrated acoustic phonon data, as shown in Fig.~2.  
The integrated magnetic scattering intensity is conserved to within error,
consistent with an accurate measurement of field effects from $H=0$ to 14~T. 
A spectrum that is symmetric about the central peak and consists of three lorentzians is fitted to the data at 14~T, as shown by the solid lines in Fig.~2.

The zero field peak is considerably broader than the energy resolution of the cold neutron spectrometer as found previously \cite{B053559}. This may be caused by a finite lifetime of the resonant spin fluctuations,
imperfect Fermi surface nesting, or broadening due to disorder on the Se/Te site.
The narrower peak width at high field (see Fig.~1) suggests another possibility where the zero field resonance encompasses overlapping anisotropy shifted components that resolve in a field.
In fitting to lower field data, we forced equal peak widths, intensity, and symmetric splitting around the central peak, as indicated by the 14 T data.  
Fig.\ 3 shows the fitted positions of spectral maxima versus field thus extracted from constant-$\bf q$ scans at several fields. To within error, the central peak position is independent of field, while the upper and lower peak positions can be described by 
\begin{equation}
\hbar\omega=\hbar\Omega_0\pm \sqrt{\delta^2+(g\mu_B B)^2},
\end{equation}
where $g=2.5(4)$ and anisotropy $\delta=1.2(4)$~meV.
Consistent with Eq.~(1), previous experiments at 7 T could not yield a peak profile substantially different from that at zero field \cite{B053559,JSWen}.

\begin{figure}

\includegraphics[width=83mm,angle=0,clip=true]{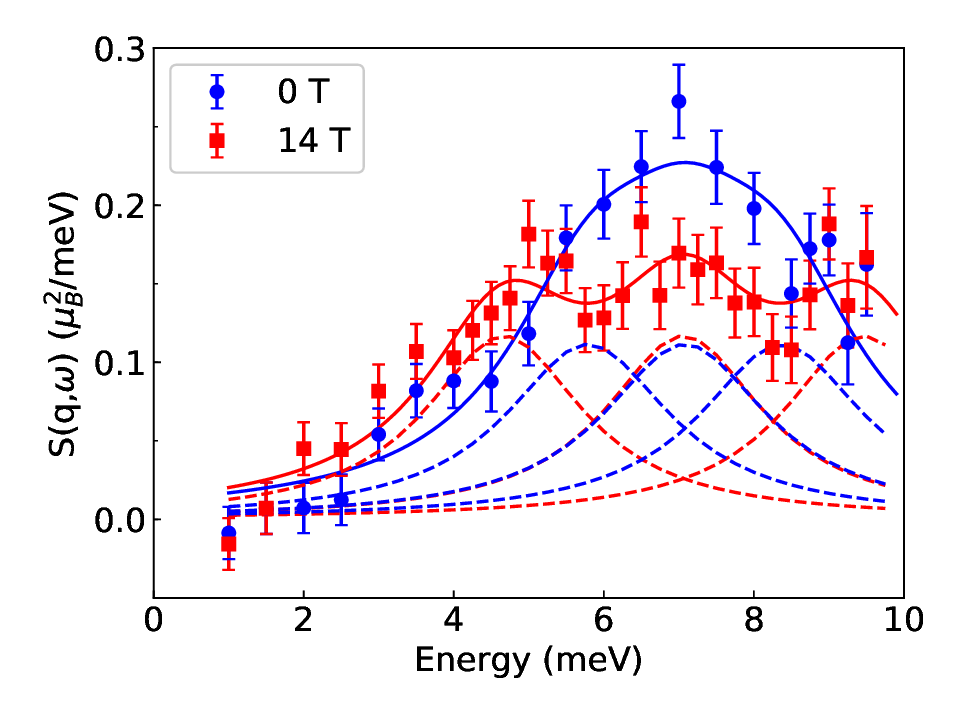}

\caption{Background subtracted low temperature inelastic scattering data from Fe$_{1.03}$Se$_{0.4}$Te$_{0.6}$ at zero field (blue circles) and 14 T (red squares). Solid lines through the data indicate fits to three lorentzians (dashed lines) constrained to have an equal width. In fitting data at fields below 14 T, we further constrained the peaks to have equal spectral weight as in the case at 14 T.}
\label{fig2}
\end{figure}

The observation of three peaks in a high magnetic field  is consistent with transitions between a singlet ground state and a triplet excited state that is field split by the Zeeman term. 
Such a level scheme can result from a near spin-space isotropic attraction between identical quasi-particles that carry spin \cite{resn_1,resn_2,resn_3,resn_t_3,A032740,A041793}. High field
extrapolation of the split peaks in Fig.~3 yields an estimate for the critical field at which the lower member of the triplet becomes degenerate with the ground state, namely $\hbar\Omega_0- \sqrt{\delta^2+(g\mu_B B_{c})^2}=0$ from Eq.~(1), therefore, 
\begin{equation}
B_{c}=\sqrt{(\hbar\Omega_0)^2-\delta^2}/g\mu_B=47(9)~T. 
\end{equation}
The remarkable match of this value and the upper critical field of $\mu_0H_{c2}\approx 47$~T inferred from high field resistivity data \cite{B095328} indicates the intimate connection between the formation of coherent singlet-triplet spectrum and superconductivity in $\rm Fe(Se,Te)$ superconductors. 

\begin{figure}

\includegraphics[width=78mm,angle=0,clip=true]{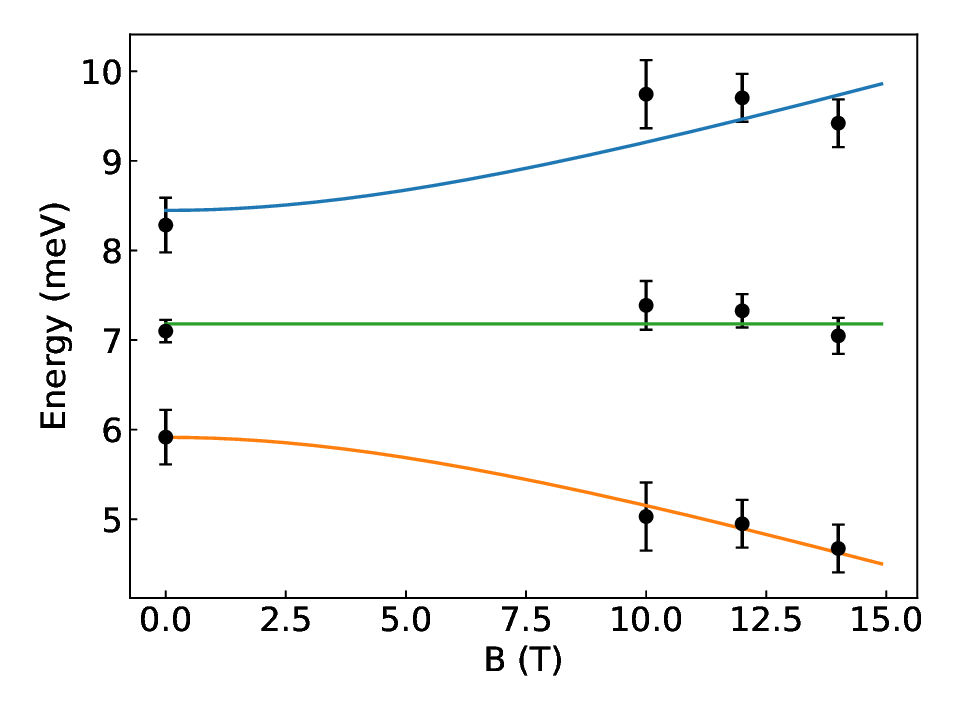}

\caption{The field dependence of peak positions extracted from the fits such as those shown in Fig.~\ref{fig2}. The solid lines through these data correspond to the fit to Eq.~(1).}
\label{fig3}
\end{figure}

We shall now address some important quantitative aspects of the high field data. With the field direction and $c$-axis as the quantization axis, the central field-independent transition must be between states with the same projection of dipole moment along the field $(\delta m=0)$ and thus the finite transition matrix element is that of the $\hat{S}_z$ operator. Conversely $\delta m=\pm 1$ for the field dependent modes, which therefore are associated with the $\hat{S}_x$ and $\hat{S}_y$ operators. Unpolarized magnetic neutron scattering probes ${\cal I}_m({\bf q},\omega)={\cal S}^{zz}+(1-\hat{q}_x^2){\cal S}^{xx}+(1-\hat{q}_y^2){\cal S}^{yy}$. Four-fold symmetry implies ${\cal S}^{xx}={\cal S}^{yy}\equiv{\cal S}^{\perp}$ so that ${\cal I}_m({\bf q},\omega)={\cal S}^{zz}+{\cal S}^{\perp}$. 
For a weakly field perturbed isotropic system where ${\cal S}^{zz}$ and ${\cal S}^{\perp}$ carry equal spectral weight, the field independent longitudinal component (the central peak) and the field dependent transverse component (the sum of the upper and lower peaks) should carry equal spectral weight. This is consistently observed for  field split singlet-triplet transitions associated with exchange coupled atomic spin degrees of freedom in a variety of insulating quantum magnets \cite{s_t_e}. In contrast, for 
the resonance in $\rm Fe_{1.03}Se_{0.4}Te_{0.6}$, the total intensity associated with the total field dependent modes is approximately twice the intensity of the field independent mode. This indicates a factor of two in enhancement of the transverse spectral density that surely is an important characteristic of the superconducting state. 

The relation between the resonance and superconducting pair correlations is a hotly debated theoretic issue \cite{so5_1998,resn_s_1999,resn_t_1,resn_t_2,resn_t_5}. A useful
diagnostic from inelastic magnetic neutron scattering  is the change in the first moment 
across the superconducting phase transition: 
\begin{equation}
\Delta \left<E({\bf q})\right>=\left(\frac{\hbar}{2\mu_B}\right)^2 \int_0^\infty \omega
{\rm Tr}\{\Delta {\cal S}({\bf q},\omega)(1-e^{-\beta\hbar\omega})\}  d\omega.
\end{equation}
For spin exchange models of magnetism, $\Delta \left<E({\bf q})\right>=-\frac{1}{3}\sum_{\bf d} J_{\bf d}\left<S_{\bf 0}\cdot S_{\bf d}\right>(1-\cos{\bf q} \cdot {\bf d})$. For an itinerant magnetic system such as $\rm Fe_{1.03}Se_{0.4}Te_{0.6}$ \cite{B114713}, we expect the amplitude 
of $\Delta \left<E({\bf q})\right>$ to monitor magnetic exchange energy through the superconducting phase transition. Assuming $\Delta {\cal S}$=${\cal S}(1.7K)-{\cal S}(30K)$ vanishes beyond a cut-off $\hbar\omega_c=13$~meV, as suggested by measurements in previous study\cite{B053559}, $\Delta \left<E({\bf q})\right>$ relative to 30 K was calculated from the normalized difference data to be $1.7(6)$~meV per Fe atom. For comparison the net condensation energy has been estimated to be $1.1$~J/mole~$\approx 0.01$ meV/Fe based on specific heat measurements\cite{B040824}. A similar orders of magnitude difference between $\Delta \left<E({\bf q})\right>$ and the net thermodynamic
condensation energy was previously noted in the heavy fermion superconductor CeCoIn$_5$ \cite{Co115_stock}. The implication is that superconductivity is driven by the reduced magnetic exchange energy that it entails. The small net condensation energy indicates an almost equal increase in the electronic kinetic energy in the superconducting state. 

More recently, a splitting of the resonance peak was also observed in heavy-fermion superconductor CeCoIn$_5$ when the magnetic field was applied in-plane, while the peak was merely broaden when the field is along the $c$-axis \cite{Co115_stock2}. In contrast to our observation in $\rm Fe_{1.03}Se_{0.4}Te_{0.6}$, however, the splitting is two-fold in the superconducting state of CeCoIn$_5$. The unexpected results of CeCoIn$_5$ have been explained to be due to the breaking of the spin-space symmetry by the strong crystalline electric field of the Ce-$4f$ quasi-particle bands \cite{Co115_theory}.
Since the Zeeman energy $\sim 2$~meV at 14 T (refer to Fig.~2) is close to the upper field limit available to current inelastic neutron scattering measurements, it is not yet feasible to apply the current experimental approach to the resonance peak of cuprate superconductors at much higher an energy.

In summary, we have presented evidence of field-induced fine structure in the spin resonance of a magnetic superconductor. Three peaks emerge from a broad zero field maximum reminiscent of Zeeman splitting of an $S=1$ triplet bound-state formed by two identical Bogoliubov-de Gennes quasi-particles with weakly anisotropic attraction. Quantitative analysis shows 
the shift in the magnetic exchange energy associated with the triplet exceeds the net condensation energy by two orders of magnitude, and the inferred critical field $B_c$ where the Zeeman energy exceeds the bound state energy matches the upper critical field. The high field neutron data show that magnetic fluctuations play a central role in iron superconductivity and suggest that the formation of a triplet bound state actually drives superconductivity in $\rm Fe_{1.03}Se_{0.4}Te_{0.6}$. 

We thank C.\ Broholm for meaningful discussions. J.L. and W.B. were supported by 
National Basic Research Program of China 
(Grant Nos. 2012CB921700 and 2011CBA00112) and the National Natural Science
Foundation of China (Grant Nos. 11034012 and 11190024). J.H. and Z.Q.M were supported 
by the NSF grant DMR-0645305 and the US DOE grant DE-FG02-07ER46358.
The work at ORNL was
supported by the U.S. Department of Energy, Office of Science, Office of Basic Energy Sciences, under contract number DE-AC05-00OR22725.

\end{document}